\documentstyle[11pt,paspconf,epsfig]{article}

\begin{document}

\title{Overview of Secondary Anisotropies of the CMB}
\author{A. Refregier}
\affil{Department of Astrophysical Sciences, Peyton Hall,
Princeton, NJ 08544}

\begin{abstract}
While the major contribution to the Cosmic Microwave Background (CMB)
anisotropies are the sought-after primordial fluctuations produced at
the surface of last scattering, other effects produce secondary
fluctuations at lower redshifts. These secondary fluctuations must be
carefully accounted for, in order to isolate primordial fluctuations.
In addition, they are interesting in their own right, since they
provide a wealth of information on the geometry and local properties
of the universe. Here, I survey the different sources of secondary
anisotropies and extragalactic foregrounds of the CMB. I show their
relative importance on the multipole-frequency plane. I discuss in
particular their impact in the future CMB missions \hbox{MAP} and
Planck Surveyor.
\end{abstract}


\section{Introduction}
The Cosmic Microwave Background (CMB) provides a unique probe of the
early universe (see White, Scott, \& Silk 1994 for a review). If CMB
fluctuations are consistent with inflationary models, future
ground-based and satellite experiments will yield accurate
measurements of most cosmological parameters (see Zaldarriaga,
Spergel, \& Seljak 1997; Bond, Efstathiou, \& Tegmark 1997 and
reference therein). These measurements rely on the detection of
primordial anisotropies produced at the surface of last
scattering. However, various secondary effects produce fluctuations at
lower redshifts. The study of these secondary fluctuations (or
extragalactic foregrounds) is important in order to isolate primordial
fluctuations.  In addition, secondary fluctuations are interesting in
their own right since they provide a wealth of information on the
local universe.

In this contribution, I present an overview of the different
extragalactic foregrounds of the CMB. The foregrounds produced by
discrete sources, the thermal Sunyaev-Zel'dovich (SZ) effect, the
Ostriker-Vishniac (OV) effect, the Integrated Sachs-Wolfe (ISW)
effect, gravitational lensing, and other effects, are briefly
described. I show their relative importance on the multipole-frequency
plane, and pay particular attention to their impact on the future CMB
missions \hbox{MAP} (Bennett et al.\ 1995) and Planck
Surveyor(Bersanelli et al.\ 1996). A more detailed account of each
extragalactic foreground can be found in the other contributions to
this volume. In this article, I have focused on the latest literature,
and have not aimed for bibliographical completeness. This overview is
based on a more detailed study of extragalactic foregrounds in the
context of the \hbox{MAP} mission (Refregier et al.  1998).

\section{Comparison of Extragalactic Foregrounds}
\label{foregrounds}
To assess the relative importance of the extragalactic foregrounds, I
decompose the temperature fluctuations of the CMB into the usual
spherical harmonic basis,
$\frac{\delta{T}}{T_{0}}(\theta)=\sum_{\ell,m} a_{lm} Y_{lm}(\theta)$,
and form the averaged multipole moments $C_{l}\equiv \langle
|a_{lm}|^{2} \rangle$. Following Tegmark \& Efstathiou (1996), I
consider the quantity $ \Delta T_{\ell} \equiv \left[ \ell(2 \ell+1)
C_{\ell}/4 \pi \right]^{\frac{1}{2}} T_{0}$, which gives the {\it rms}
temperature fluctuations per $\ln \ell$ interval centered at
$\ell$. Another useful quantity that they considered is the value of
$\ell=\ell_{eq}$ for which foreground fluctuations are equal to the
CMB fluctuations, i.e.\ for which $C_{\ell}^{\rm foreground} \simeq
C_{\ell}^{\rm CMB}$. Note that, since the foregrounds do not
necessarily have a thermal spectrum, $\Delta T_{\ell}$ and $\ell_{eq}$
generally depend on frequency.

\begin{figure}
\centerline{\epsfig{file=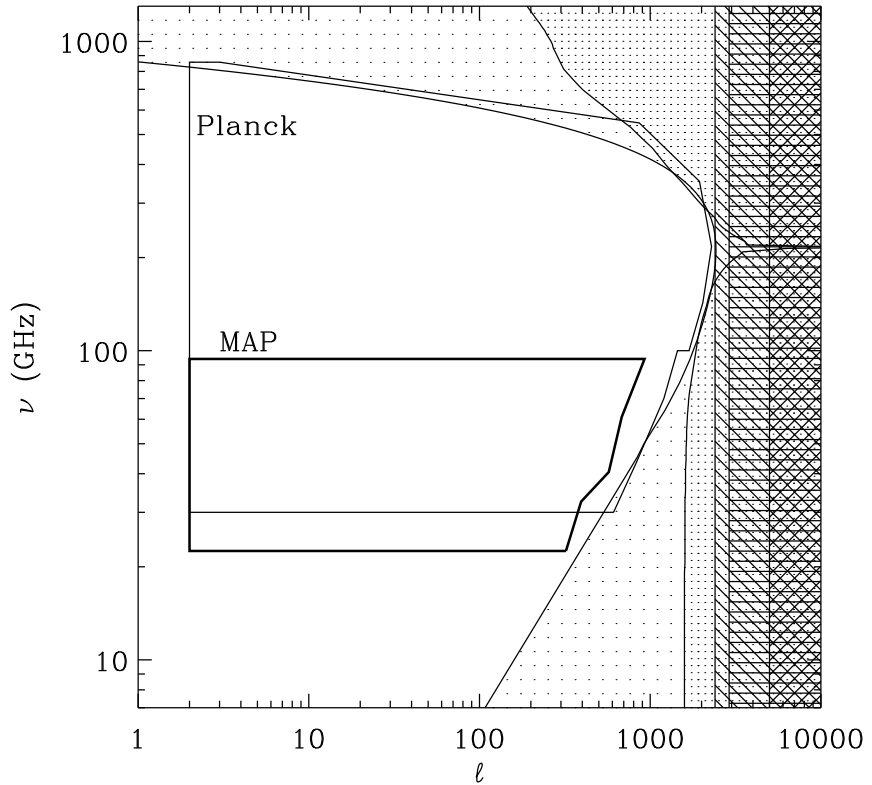,width=12cm}}
\caption{Summary of the importance of extragalactic foregrounds of the
CMB. Each filled area on the multipole-frequency plane corresponds to
regions where the foreground fluctuations exceed those of the CMB. The
sparse and dense dotted regions correspond to discrete sources (with
$S<1$ Jy) and the Sunyaev-Zel'dovich effect (with cluster cores),
respectively. The horizontal, descending and ascending hashed regions
correspond to the Ostriker-Vishniac effect (with $z_{r}=50$),
gravitational lensing, and the integrated Sachs-Wolfe effect (with
$\Omega h = 0.25$), respectively. The areas marked \hbox{MAP} (thick
line) and Planck (thin line) show the regions of sensitivity for each
of the future missions, i.e.\ regions where the CMB fluctuations
exceed the noise for each of the instruments (see text).}
\label{fig:lnu}
\end{figure}

The comparison is summarized in table~\ref{tab:foregrounds} and in
figure~\ref{fig:lnu}. Table~\ref{tab:foregrounds} shows $\Delta
T_{\ell}$ and $\ell_{eq}$ for each of the major extragalactic
foregrounds at $\nu=94$ GHz and $\ell=450$, which corresponds to a
FWHM angular scale of about $\theta \sim .3$ deg. These values were
chosen to be relevant to the \hbox{MAP} W-band ($\nu \simeq 94$ GHz
and $\theta_{\rm beam} \simeq 0\fdg21$. I also indicate whether each
foreground component has a thermal spectrum.

Figure~\ref{fig:lnu} summarizes the importance of each of the
extragalactic foregrounds in the multipole-frequency plane. It should
be compared to the analogous plot for galactic foregrounds (and
discrete sources) shown in Tegmark \& Efstathiou (1996; see also
Tegmark 1997 for an updated version). These figures show regions on
the $\ell$-$\nu$ plane in which the foreground fluctuations exceed the
CMB fluctuations, i.e.\ in which $C_{\ell}^{\rm foreground} >
C_{\ell}^{\rm CMB}$.  As a reference for $C_{\ell}^{\rm CMB}$, a COBE
normalized CDM model with $\Omega_{b}=0.05$ and $h=0.5$ was used. Also
shown in figure~\ref{fig:lnu} is the region in which \hbox{MAP} and
Planck Surveyor are sensitive, i.e.\ in which $\Delta C_{\ell}^{\rm
noise} < C_{\ell}^{\rm CMB}$, where $\Delta C_{\ell}^{\rm noise}$ is
the {\it rms} uncertainty for the instrument. Note that this figure is
only intended to illustrate the domains of importance of the different
foregrounds qualitatively.

In the following, I briefly describe each extragalactic foreground
and comment on its respective entries in table~\ref{tab:foregrounds}
and figure~\ref{fig:lnu}.

\begin{table}
\caption{Summary of Extragalactic Foregrounds for $\nu=94$
GHz and $\ell=450$.}
\label{tab:foregrounds}
\begin{center}\small
\begin{tabular}{rrrrrr}
Source & $\Delta T_{\ell}$ ($\mu$K)\tablenotemark{a} &
  $\ell_{\rm eq}$\tablenotemark{b} & Thermal\tablenotemark{c} &
  Note & Ref.\tablenotemark{d} \\
\tableline
CMB\tablenotemark{e} & 50 & & yes & & 1\\
Discrete\tablenotemark{f} & 5   & 1800      & no & $S<1.0$  Jy  & 2 \\
                          & 2   & 3100      & no & $S<0.1$ Jy     & 2 \\
SZ\tablenotemark{g} & 10   & 1900      & no & C     &3 \\
                    & 7    & 2300      & no & NC    &3 \\ 
OV\tablenotemark{h} & 2   & 2900      & yes & $z_{r}=50$    & 4 \\
                    & 1   & 3100      & yes & $z_{r}=10$    & 4 \\
ISW & 1     & 5000      & yes & $\Omega h=0.25$     & 5 \\
    & 0.9   & 5800      & yes & $\Omega h=0.50$      & 5 \\
Lensing & 5 & 2400 & yes & & 6
\end{tabular}
\end{center}
%\normalsize
\tablenotetext{a}{$\Delta T_{\ell} \equiv [\ell(2\ell +1)C_{\ell}/4\pi]^{1/2}$
centered at $\ell=450$ and $\nu=94$ GHz.}
\tablenotetext{b}{Value of $\ell$ for which $\Delta T_{\ell}=\Delta
T_{\ell,{\rm CMB}}$}
\tablenotetext{c}{Thermal (yes) or nonthermal (no) spectral dependence}
\tablenotetext{d}{1: Seljak \& Zaldarriaga 1996; 2: Toffolatti
et al.\ 1998; 3: Persi et al.\ 1995; 4: Hu \& White 1996;
5: Seljak 1996a; 6: Zaldarriaga \& Seljak 1998a}
\tablenotetext{e}{Primordial CMB fluctuations for a CDM model
with $\Omega_{m}=1$, $\Omega_b=0.05$, and $h=0.5$}
\tablenotetext{f}{Discrete sources with 94 GHz removal threshold of
0.1, 1 Jy, respectively}
\tablenotetext{g}{SZ effect with (C) and without (NC) cluster cores
respectively.}
\tablenotetext{h}{OV effect with two different reionization redshifts
$z_{r}$}
\end{table}

\subsection{Discrete Sources}
Discrete sources produce positive, point-like, non-thermal
fluctuations. While not much is known about discrete source counts
around $\nu \sim 100$ GHz, several models have been constructed by
interpolating between radio and IR observations
(Toffolatti et al.\ 1998; Gawiser \& Smoot 1997; Gawiser et al. 1998;
Sokasian et al. 1998). Here, I adopt the model of
Toffolatti et al.\ and consider the two flux limits $S <1$ and 0.1 Jy
for the source removal in table~\ref{tab:foregrounds}.  The sparsely
dotted region figure~\ref{fig:lnu} shows the discrete source region
for $S <1$ Jy.  In the context of CMB experiments, the Poisson shot
noise dominates over clustering for discrete sources (see Toffolatti
et al.\ ). As a result, the discrete source power spectrum,
$C_{\ell}^{\rm discrete}$, is essentially independent of
$\ell$.

\subsection{Thermal Sunyaev-Zel'dovich Effect}
\label{foregrounds_sz}
The hot gas in clusters and superclusters of galaxies affect the
spectrum of the CMB through inverse Compton scattering. This effect,
known as the Sunyaev-Zel'dovich effect (for reviews see Sunyaev \&
Zel'dovich 1980; Rephaeli 1995), results from both the thermal and
bulk motion of the gas. We first consider the thermal SZ effect, which
typically has a larger amplitude and has a non-thermal spectrum (see
the \S\ref{OV} below for a discussion of the kinetic SZ effect). The
CMB fluctuations produced by the thermal SZ effect have been studied
using the Press-Schechter formalism (see Bartlett 1997 for a review),
and on large scales using numerical simulations (Cen \& Ostriker 1992;
Scaramella, Cen, \& Ostriker 1993) and semi-analytical methods (Persi
et al.\ 1995). Here, I consider the SZ power spectrum, $C_{\ell}^{\rm
SZ}$, calculated by Persi et al.\ (see their figure 5). In
table~\ref{tab:foregrounds}, I consider their calculation both with
and without bright cluster removal. In figure~\ref{fig:lnu}, only the
spectrum without cluster removal is shown.

\subsection{Ostriker-Vishniac Effect}
\label{OV}
In addition to the thermal SZ effect described above, the hot
intergalactic medium can produce thermal CMB fluctuations as a result
of its bulk motion. While this effect essentially vanishes to first
order, the second order term in perturbation theory, the
Ostriker-Vishniac effect (Ostriker \& Vishniac 1986; Vishniac 1987),
can be significant on small angular scales. The power spectrum of the
OV effect depends on the ionization history of the universe, and has
been calculated by Hu \& White (1996), and Jaffe \& Kamionkowski
(1998; see also Persi et al.\ 1995). We use the results of Hu \& White
(see their figure 5) who assumed that the universe was fully reionized
beyond a redshift $z_{r}$.  In table~\ref{tab:foregrounds}, I consider
the two values $z_{r}=10$ and 50, while in figure~\ref{fig:lnu}, I
only plot the region corresponding to $z_{r}=50$. For consistency, the
standard CDM power spectrum is still used as a reference, even though
the primordial power spectrum would be damped in the event of early
reionization. (Using the damped primordial spectrum makes, at any
rate, only small corrections to both table~\ref{tab:foregrounds} and
figure~\ref{fig:lnu}.)

\subsection{Integrated Sachs-Wolfe Effect}
The Integrated Sachs-Wolfe Effect (ISW) describes thermal CMB
fluctuations produced by time variations of the gravitational
potential along the photon path (Sachs \& Wolfe
1967). Linear density perturbations produce non-zero ISW fluctuations
in a $\Omega_m \neq 1$ universe only. Non-linear perturbations produce
fluctuations for any geometry, an effect often called the Rees-Sciama
effect (Rees \& Sciama 1968). Tuluie
\& Laguna (1995) have shown that anisotropies due to intrinsic changes
in the gravitational potentials of the inhomogeneities and
anisotropies generated by the bulk motion of the structures
across the sky generate CMB anisotropies in the range of
$10^{-7} \la \frac{\Delta T}{T} \la 10^{-6}$ on scales of about
$1^{\circ}$ (see also Tuluie et al. 1996). The power
spectrum of the ISW effect in a CDM universe was computed by
Seljak (1996a; see also references therein). In
table~\ref{tab:foregrounds}, I consider values of the density
parameter, namely $\Omega h=0.25$ and $0.5$. In figure~\ref{fig:lnu},
only the $\Omega h=0.25$ case is shown. As above, the standard CDM
($\Omega =1$, $h=0.5$) spectrum is still used as a reference.

\subsection{Gravitational Lensing}
Gravitational lensing is produced by spatial perturbations in the
gravitational potential along the line of sight (see Schneider,
Ehlers, \& Falco 1992; Narayan \& Bartelmann 1996). This effect does
not directly generate CMB fluctuations, but modifies existing
background fluctuations. The effect of lensing on the CMB power
spectrum was calculated by Seljak (1996b) and Metcalf \& Silk
(1997). Recently, Zaldarriaga \& Seljak (1998a) included the lensing
effect in their CMB spectrum code (CMBFAST; Seljak \& Zaldarriaga
1996). This code was used to compute the absolute lensing correction
$|\Delta C_{\ell}^{\rm lens}|$ to the standard CDM spectrum, including
nonlinear evolution. The results are shown in
table~\ref{tab:foregrounds} and figure~\ref{fig:lnu}.

\subsection{Other Extragalactic Foregrounds}
In addition to the effects discussed above, other extragalactic
foregrounds can cause secondary anisotropies.  For instance, patchy
reionization produced by the first generation of stars or quasars can
cause second order CMB fluctuations through the doppler effect
(Aghanim et al. 1996a,b; Gruzinov \& Hu 1998; Knox, Scoccimaro,
\& Dodelson 1998; Peebles \& Juzkiewicz 1998).
Calculations of the spectrum of this effect are highly uncertain, but
show that the resulting CMB fluctuations could be of the order of 1
$\mu$K on 10 arcminute scales, for extreme patchiness. More likely
patchiness parameters make the effect negligible on these scales, but
potentially important on arcminute scales. Another potential
extragalactic foreground is that produced by the kinetic SZ effect
from Ly$_{\alpha}$ absorption systems, as was recently proposed by
Loeb (1996). The resulting CMB fluctuations are of the
order of a few $\mu$K on arcminute scales, and about one order of
magnitude lower on 10 arcminute scales.  Because of the uncertainties
in the models for these two foregrounds and because they are small on
10 arcminute scales, they are not included in
table~\ref{tab:foregrounds} and figure~\ref{fig:lnu}.

\section{Discussion and Conclusion}
An inspection of table~\ref{tab:foregrounds} shows that, at 94 GHz and
$\ell=450$, the power spectra of the largest extragalactic foregrounds
considered are a factor of 5 below the primordial CDM spectrum. As can
be seen in figure~\ref{fig:lnu}, the dominant foregrounds for
\hbox{MAP} and Planck Surveyor are discrete sources, the thermal SZ
effect and gravitational lensing. Note that, for Planck surveyor,
these three effects produce fluctuations which are close to the
sensitivity of the instrument. The spectra of the OV and ISW effects
will produce fluctuations of the order of $1 \mu$K , and are thus less
important for a measurement of the power spectrum.  The effect of
gravitational lensing is now incorporated in CMB codes such as
CMBFAST, and can thus be taken into account in the estimation of
cosmological parameters. The other two dominant extragalactic
contributions, discrete sources and the thermal SZ effect, must also
be accounted for, but are more difficult to model. Note that, on large
angular scales, extragalactic foregrounds produce relatively small
fluctuations, and are thus not detectable in the COBE maps (Boughn \&
Jahoda 1993; Bennett et al.\ 1993; Banday et al.\ 1996; Kneissl et
al.\ 1997)

While I have concentrated above on the power spectrum, secondary
anisotropies are also a source of non-gaussianity in CMB
maps. Discrete sources and the SZ effect from clusters of galaxies
mainly produce Poisson fluctuations and are thus clearly
non-gaussian. The other extragalactic foregrounds (SZ, OV, ISW, and
lensing) are also non-gaussian and trace large-scale structures in the
local universe. As a consequence of the latter fact, extragalactic
foregrounds can be probed by cross-correlating CMB maps with galaxy
catalogs, which act as tracers of the large scale structure.  Such
technique can be used to detect the ISW effect (Boughn et al. 1998,
and reference therein), gravitational lensing (Suginohara et al. 1998)
and the SZ effect by superclusters (Refregier et
al. 1998). Gravitational lensing is particularly interesting since it
produces a specific non-gaussian signature (Bernardeau 1998). This
signature can be used to reconstruct the gravitational potential
projected along the line of sight (Zaldarriaga \& Seljak 1998b).
Further non-gaussian signatures result from the fact that the
different extragalactic foregrounds are spatially correlated. For
instance, a detection of the cross-correlation signal between
gravitational lensing and the ISW and SZ effects would allow us to
determine the fraction of the ionized gas and the time evolution of
gravitational potential (Goldberg \& spergel, 1998; Seljak \&
Zaldarriaga 1998).  A detection of secondary anisotropies would help
break the degeneracy between cosmological parameters measured from
primary anisotropies alone.

\acknowledgments I thank David Spergel and Thomas Herbig for active
collaboration and discussions on this project.  This work was
supported by the \hbox{MAP} MIDEX program.

\end{document}